\documentclass[journal]{IEEEtran}

\makeatletter
\def\ps@headings{%
\def\@oddhead{\mbox{}\scriptsize\rightmark \hfil \thepage}%
\def\@evenhead{\scriptsize\thepage \hfil \leftmark\mbox{}}%
\def\@oddfoot{}%
\def\@evenfoot{}}
\makeatother
\usepackage{algorithm}
\usepackage{algorithmic}
\usepackage{amsthm}
\usepackage{mathtools}
\usepackage{amssymb}
\usepackage{graphicx}
\usepackage{cite}
\usepackage{amssymb, amsmath, amsfonts, epsfig, latexsym, times}
\usepackage{bm}
\usepackage{subfigure}
\usepackage{balance}

\begin{document}

\title{Security Schemes in Vehicular Ad hoc Networks with Cognitive Radios}
\author{\IEEEauthorblockN{Zhexiong~Wei\IEEEauthorrefmark{1}, F.~Richard~Yu\IEEEauthorrefmark{1}, Helen Tang\IEEEauthorrefmark{2}, Chengchao Liang\IEEEauthorrefmark{1}, and Qiao Yan\IEEEauthorrefmark{3}}

\IEEEauthorblockA{\IEEEauthorrefmark{1}Depart. of Systems and Computer Eng., Carleton University, Ottawa, ON, Canada}

\IEEEauthorblockA{\IEEEauthorrefmark{2}Defense R\&D Canada, Ottawa, ON, Canada}

\IEEEauthorblockA{\IEEEauthorrefmark{3}College of Computer Science and Software Engineering, \\Shenzhen University,  Shenzhen, Guangdong, China}\\
Email: zhexiong\_wei@carleton.ca, richard.yu@carleton.ca, helen.tang@drdc-rddc.gc.ca, \\
chengchaoliang@sce.carleton.ca, yanq@szu.edu.cn
}


\maketitle

\begin{abstract}
Vehicular Ad hoc NETworks (VANETs) as the basic infrastructure can facilitate applications and services of connected vehicles (CVs). Cognitive radio (CR) technology is an effective supplement and enhancement for VANETs. It can reduce the impact of  deficiency of spectrum resource in VANETs. Although CR-VANETs can utilize the unused licensed spectrum effectively, the distributed nature of CR-VANETs may open a door for different attacks, such as spectrum sensing data falsification attack. In this paper, we propose a joint RSU and vehicle-based light-weighted cloud for CR-VANETs. Based on this cloud computing model, we propose a new service named Spectrum Sensing as a Service (SSaaS), which can perform a cooperative spectrum sensing in CR-VANETs with cloud computing assistance to secure the spectrum sensing procedure. As a result, a reliable service can be obtained in CR-VANETs. Simulation results show that the cloud computing in CR-VANETs can effectively reduce latency and improve the security of CR-VANETs.

\end{abstract}
\begin{IEEEkeywords}
Cognitive radio, vehicular ad hoc networks, cloud computing, security.
\end{IEEEkeywords}

\section{Introduction}
Recently, there is a phenomenal burst of interest in connected vehicles (CVs). CV systems use connectivity (via advanced wireless communications) to enable vehicles, smart roadway infrastructure (SRI) and personal mobile devices to exchange information with each other, and to provide road users with both safety and mobility advisories, warnings and alerts \cite{Yu2016}. Vehicular Ad hoc NETworks (VANETs) as the basic infrastructure can facilitate applications and services of CVs \cite{BLB16}. Safety and entertainment are the main topics and incentives of CVs, which are drawing great interest from both academia and industry. To achieve these two primary goals, several challenges in VANETs need to be solved. The first one is the \emph{limited resources}, such as radio spectrum resources, computation and storage resources. The second one is the \emph{latency requirement} from driving applications and services, which is stricter than traditional network applications' requirements. In addition, \emph{security} is also very critical for VANETS.

For the limited radio spectrum issue, a  promising solution is \emph{cognitive radio} \cite{S05,Yu11}, which has already extended to VANETs environments \cite{S05}. Generally, CR technology can help secondary users (SUs) that have no pre-defined spectrum for wireless communications to utilize the primary users (PUs)' licensed spectrum. The condition is that the SUs should not interrupt the PUs' normal communications. In other words, SUs should have the ability to detect the presence of PU and concede the licensed spectrum for the PU. In addition, PUs have no requirement to change their existing infrastructure to accommodate SUs \cite{LYJ10_ICC,ALVM06,WTY09,Yu11,YHT10,ML13,RNM12,BYL11_TWC,LGL14,BYL11_TVT}. In VANETs, CR technology can be used to mitigate the spectrum limitation issue.

For the issue of limited computation and storage resources, cloud computing in VANETs can leverage the limited capability of a specific vehicle by using the idle computing units and storage space from other vehicles and roadside units (RSUs) \cite{BMZ2015}. Cloud computing for VANETs can basically provide three types of services for the vehicle tenant: software as a service (SaaS), platform as a service (PaaS) and infrastructure as a service (IaaS) \cite{BMZ2015}.  For SaaS, the vehicle can perform the applications or services, which are already in the cloud computing service, e.g., weather forecasting. For PaaS, the vehicle can perform its applications on the platform, such as operating system, provided by cloud services. For IaaS, the vehicle can request a larger space on the cloud to store the entertainment videos temporarily.

The low latency requirement is extremely important in CR-VANETs with cloud computing due to the high dynamic characteristics in the driving scenarios \cite{SW2014}. This makes the traditional and powerful cloud on the  Internet hard to satisfy the applications and services in CR-VANETs. The vehicular cloud, which is comprised of vehicles and RSUs autonomously, is an attractive and possible alternative. Compared to the traditional cloud, it can have lower latency \cite{HLC16}.

Last but not least, security in the VANETs is the key to make any service or application successful in the safety driving condition. Security permeates every corner in CR-VANETs, from spectrum sensing to networking, from software to cloud computing. In spectrum sensing, the sensing data falsification attack (SSDF) \cite{CPK08} is  a notable attack. In networking, the packet dropping attack or black hole attack is a big threat\cite{LLO11}. These security issues need to be carefully addressed.

Although several research works have been done to address  security issues in CR-VANETs and cloud computing in VANETs, securing CR-VANETs with trusted light-weighted cloud computing has not been studied yet. In this paper, we propose a joint RSU and vehicle-based light-weighted cloud for CR-VANETs. Based on this cloud computing model, we propose a new service named Spectrum Sensing as a Service (SSaaS), which can perform a cooperative spectrum sensing in CR-VANETs with cloud computing assistance to secure the spectrum sensing procedure. As a result, a reliable service can be obtained in CR-VANETs. Simulation results show that the cloud computing in CR-VANETs can effectively reduce latency and improve the security of CR-VANETs.

The remainder of this paper is outlined as follows. Related work is presented in Section \ref{sect:related_work}. We describe the CR-VANETs in Section \ref{sect:cr_vanets}. The cloud computing model in VANETs is introduced in Section \ref{sect:cloud_computing}. Then we explain the security issues in CR-VANETs in Section \ref{sect:security_issue}. Next, the proposed architecture is illustrated in Section \ref{sect:proposed_scheme} The simulation results are showed and discussed in Section \ref{sect:simulation}. Finally, we give the conclusion of the work in Section \ref{sect:dso}.

\section{Related Work}
\label{sect:related_work}

Security and QoS provisioning are two important issues in networks \cite{MYL04, YL01, LYH10, YK07, XYJL12, ATV12, YWL06_MONET,LY15, WYS10, GYJ10, BYC12, XYJ12, YTH09, YHT10, LYJ10, YKL06,YZX11, LYJ15,GYJ11,BY14,LYL09,YYG15,BYY15,ZYN12_JSAC,ZYL10,WTY14,BYL11_Online,BY13,YTM10}. Security issues of spectrum sensing in CR are studied extensively, and several mitigation schemes are proposed in the literature \cite{CPK08,KKB10,WLSH09}. A Bayesian methodology is used in\cite{WLSH09} to detect and filter out the malicious nodes with a threshold. The researchers in \cite{CPK08} present a  Weighted Sequential Probability Ratio Test to protect spectrum sensing from SSDF attacks. The authors of \cite{KKB10} utilize outlier detection techniques to identify malicious SUs. The techniques don't require full information of PU and the amount of sensing data samples is less than in other schemes. A distributed cooperative spectrum sensing mechanism is studied in \cite{LYH10}. The authors describe a consensus-based algorithm to detect the status of PU. A variant SSDF attack, named covert adaptive data injection attack is proposed in \cite{YLJLH12}, which can attack the node during the distributed consensus-based spectrum sensing.

Several cloud computing models in VANETs are proposed. In \cite{BMZ2015}, the authors propose a cloud computing model for VANETs, named VANET-Cloud, which extends the cloud to the vehicles. In this model, there are two types: permanent and temporary. A vehicular cloud networking model is proposed in \cite{LLGO2014}, which combines the vehicular cloud computing and information-centric networking. This model can provide efficient services for drivers. The authors in \cite{YZW2016} present a vehicular social network in cloud computing, which can help the useful information transmission for users who are interested in. In \cite{YZGXY2013}, a three-layer cloud computing model in VANETs is presented. In this cloud-based vehicular network, the resource management is studied with a game-theoretical mechanism. Resource management for cognitive cloud vehicular networks is studied in \cite{CAB2015}. The authors use an optimization scheme to solve this resource management problem. Fog computing paradigm in VANETs is studied in \cite{SW2014}. Due to the requirement of dynamic, local, and delay-limited nature of VANETs, fog computing can become a useful alternative of conventional cloud computing. In \cite{BCSCADPV2016}, the authors propose vehicular node virtualization, which allows the service provider to allocate resources to tenants within vehicular nodes. The virtual machine migration in VANETs is also studied.

\section{Security Issues in Cognitive Radio Vehicular Ad Hoc Networks with Cloud Computing}

In this section, we first introduce CR-VANETs, followed by cloud computing in CR-VANETs. Then, the security issues are discussed.
\subsection {Cognitive Radio Vehicular Ad Hoc Networks}
\label{sect:cr_vanets}
\begin{figure}[tp]
\centering
\includegraphics[width=0.45\textwidth]{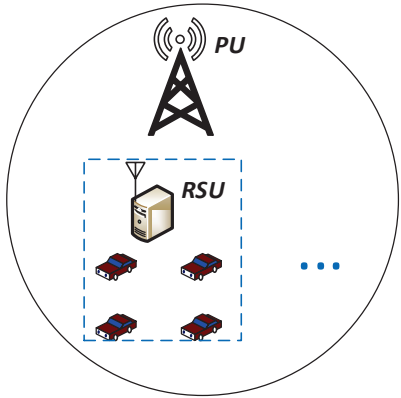}
\caption{A cognitive radio vehicular ad hoc network with RSU.}
\label{fig1}
\end{figure}

CR-VANETs are proposed to solve the issue of spectrum shortage in vehicular networks. Vehicles equipped with CR technology can communicate with each other through the licensed spectrums owned by the PU. These vehicles form a network, which is the secondary network. Cooperative spectrum sensing can be adopted in  CR-VANETs due to its dynamic and mobile nature. Each vehicle detects the presence of the PU independently. Energy detection of spectrum sensing can be used as the detection method of each vehicle due to its simplicity. In this paper, we consider the RSUs as fixed units, which  can also participate the cooperative spectrum sensing process to improve the accuracy of the sensing results. Fig.\ref{fig1} shows a CR-VANET that consists of several vehicles and a RSU. SUs (including vehicles and RSU) in this network can perform cooperative spectrum sensing in order to detect the status (i.e., presence or absence) of the PU.

\subsection{Cloud Computing in VANETs}
\label{sect:cloud_computing}
Cloud computing in VANETs can extend the computation ability and storage space of each vehicle. There are three types of services that cloud computing in VANETs can provide: software as a service, infrastructure as a service, and platform as a service. Due to the mobility and constraints of cloud computing in VANETs, the cloud is typically categorized by three levels hierarchically: Internet-based cloud, RSU-based cloud, and vehicle-based cloud. Internet-based cloud is the conventional cloud, which has amounts of computing ability, storage space and bandwidth. But it has a large latency. RSU-based cloud is the local cloud, which is close to the end user. It has limited resource of computation, storage and bandwidth, compared to the Internet-based cloud. But it has a lower latency. Vehicle-based cloud is a temporary cloud, which consists of vehicles. Vehicle-based cloud  plays a dual role: service provider and service consumer. Thus this cloud has very limited resource and the lowest latency.

\begin{figure}[tp]
\centering
\includegraphics[width=0.48\textwidth]{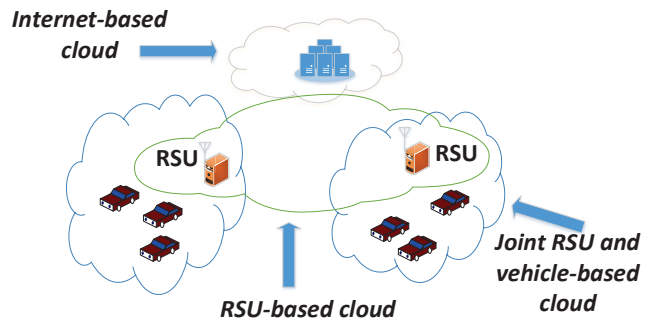}
\caption{Joint RSU and vehicle-based cloud in CR-VANETs.}
\label{fig2}
\end{figure}

In this paper, we adopt a hybrid local cloud computing model, named joint RSU and vehicle-based cloud, which combines the RSU-based cloud and vehicle-based cloud. This model can bring the fixed RSU resource to the mobile vehicular resource. It reduces the impact of unstable nature from vehicles and latency of the RSU cloud. The existing three layers cloud model can still provide services in the high level. Fig. \ref{fig2} shows that the Internet-based cloud is the highest level cloud. RSU cloud is in the middle level. The lowest level is the joint RSU and vehicle-based cloud.

Based on the joint RSU and vehicle-based cloud, we propose a new service named Spectrum Sensing as a Service (SSaaS), which can virtualize the vehicle in CR-VANETs to perform cooperative spectrum sensing. There are three distinguished advantages for vehicular virtualization in CR-VANETs as follows. 1). Vehicles that need to perform cooperative spectrum sensing can obtain more computation and storage resources. 2). Vehicles can join the spectrum sensing without contacting with other physical vehicles. 3). SSaaS can automatically and ubiquitously find other vehicles when the specific vehicle is not available.

\subsection{Security Issues in CR-VANETs}
\label{sect:security_issue}
Security issues have been studied in spectrum sensing for several years. The incumbent emulation (IE) attack is proposed in \cite{CPR2008}. In this attack, a malicious CR-enabled node mimics the PU's signal characteristics in order to interrupt the spectrum sensing process.
The SSDF attack\cite{CPH08} is the most famous one, in which malicious SUs deliberately disseminate wrong sensing data to others so that the cooperative spectrum sensing is destroyed.
For CR-VANETs, the SSDF is harder to be mitigated due to the mobility of each vehicle and limited resource for the protection of crypto-systems, such as public key infrastructure (PKI) based mechanisms.

In addition to the malicious attacks in the spectrum sensing phase, traditional security threats are still concerns in CR-VANETs, such as packet dropping attacks, also known as black hole attacks \cite{LLO11}. In this attack, the vehicle in the middle of the source and destination nodes can drop any packets, which need to be forwarded, including control packets and data packets.

\section{Securing CR-VANETs with Trusted Light-weight Cloud Computing}
\label{sect:proposed_scheme}
Conventional consensus-based spectrum sensing requires physical nodes, such as vehicles, to work together. In CR-VANETs, during the spectrum sensing process, each vehicle has to stay in the local network. Another drawback is the limited resource of each physical vehicle. Each vehicle has different capability, e.g., computation, storage and bandwidth, to perform the spectrum sensing algorithm. To solve these issues, we present a cloud-based secure spectrum sensing for CR-VANETs, named SSaaS. This service is supported by the jointed RSU and vehicle-based cloud, which is described in Subsection \ref{sect:cloud_computing}. Firstly, we introduce the architecture of the SSaaS. Then we explain the general service deployment template for the SSaaS. Finally, the consensus-based spectrum sensing with trust is illustrated as a service deployment in the SSaaS.

\begin{figure}[tp]
\centering{
\includegraphics[width=0.48\textwidth]{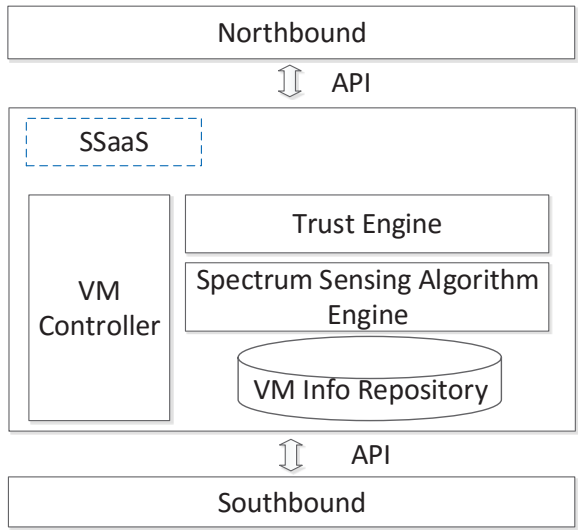}
\caption{The architecture of SSaaS.}
\label{pic:architecture_ssaas}}
\end{figure}

\subsection{Architecture of SSaaS}
In order to leverage spectrum sensing in cloud computing, an architecture of SSaaS is presented in Fig. \ref{pic:architecture_ssaas}. the Northbound could be a user or application, which is interested in a service deployment in the SSaaS. The Southbound could be any basic cloud computing platform such as OpenStack \cite{OPENSTACK}. The SSaaS module provides core functions for spectrum sensing in CR-VANETs. The API between these three modules can be REST-enabled, which is popular in the cloud computing development community, or traditional library, which is embedded in the upper application software. In the SSaaS module, the VM controller orchestrates all the VMs in the service based on the service description in the template. For example, the VM locates in the specific host or which spectrum sensing algorithm is activated. The trust engine module provides a variety of trust mechanisms, which are used to assess the trustworthiness \cite{WTY14} of each vehicle. The spectrum sensing algorithm engine is in charge of the specific spectrum sensing, which is selected in the service deployment. The  VM info repository stores the specific service deployment details.

\subsection{SSaaS Service Deployment Template}
\begin{figure}[tp]
\centering{
\includegraphics[width=0.65\textwidth]{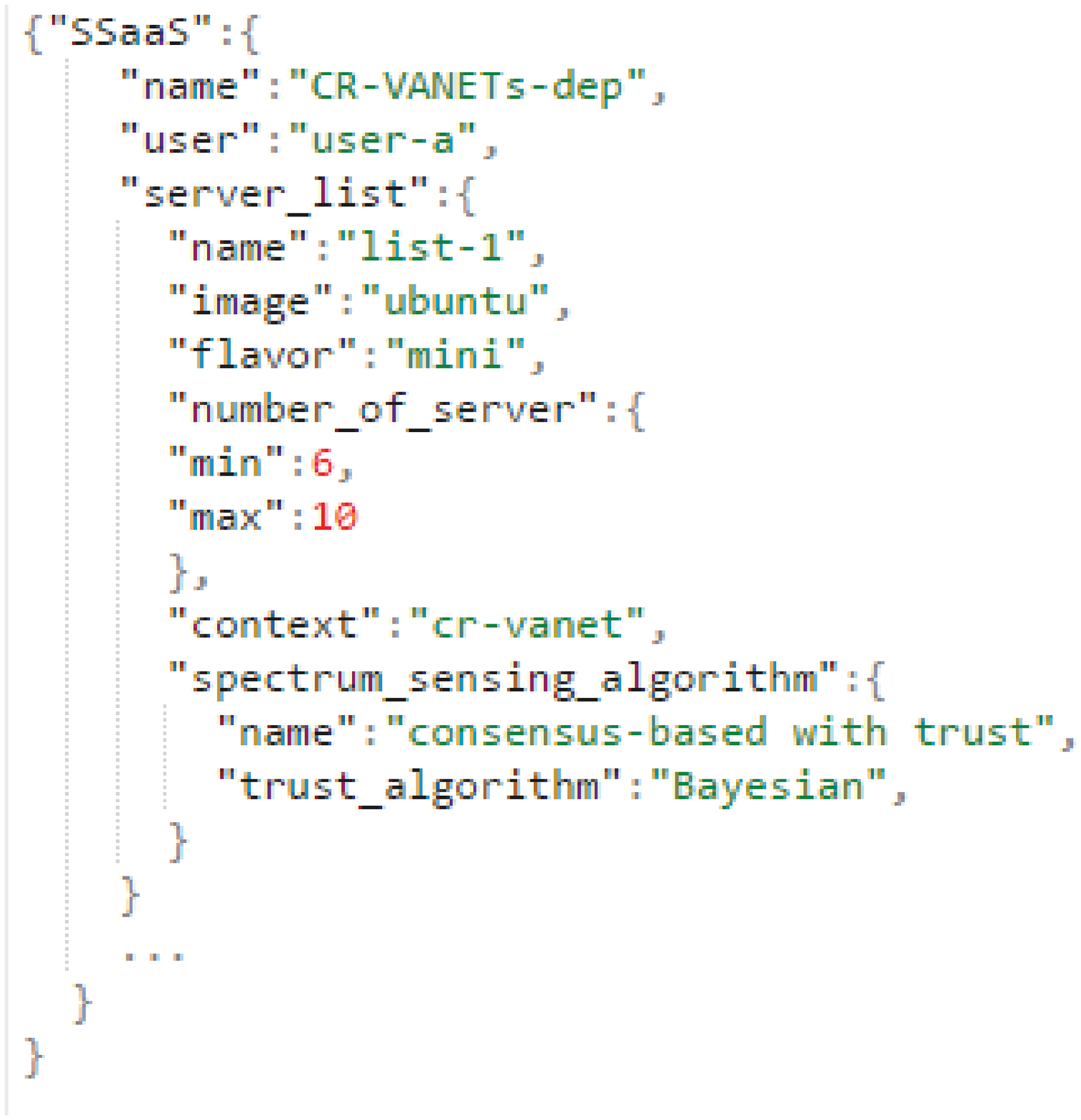}
\caption{SSaaS service deployment template in JSON.}
\label{pic:dep_template}}
\end{figure}

A user or application that needs to use SSaaS can define a detailed service deployment with a template, which is similar to the conventional cloud service deployments \cite{HEAT}. We format the template with JSON \cite{JSON}, which is not only processed by most modern programming languages but also friendly for human-reading. Fig. \ref{pic:dep_template} shows an SSaaS deployment from Northbound. In this service deployment, the number of servers is defined that how many VMs join together to perform cooperative spectrum sensing. Context is used to describe the network environment. The spectrum sensing algorithm specifies the scheme for spectrum sensing in the service. The trust algorithm defines the trust scheme that trust engine will processes. We will explain it with more details in the next subsection.

\subsection{Consensus-based Spectrum Sensing with Trust}
There are two important parts in this scheme. Firstly, the consensus-based spectrum sensing is introduced. Secondly, trust scheme can assist the consensus-based spectrum sensing, where trust is considered as the weight.

Here we formulate a CR-VANET as an undirect simple graph, $G$. This graph is represented by a matrix $A = (a_{ij})_{n\times n}$, where
\begin{equation}
a_{ij}=\left\{\begin{aligned}
         \begin{array}{ll}
          1, & \mbox{if}\  j \in N(i) \\
          0, & \mbox{otherwise}
         \end{array}
         \end{aligned} \right.
\end{equation}
where $N(i)$ is a one-hop neighbor set of vertex $i$.

The weighted-average consensus \cite{T08,M05} is listed as follows:
\begin{equation}
\label{weighted_consensus}
x_i(k+1)=x_i(k)+\epsilon \sum_{j\in N_i(k)}w_{ij}(x_j(k)-x_i(k)),
\end{equation}
where $0< \epsilon <(\smash{\displaystyle\max_i |N_i|})^{-1}$, $w_{ij}$ is a weight for CR vehicle $i$ to its neighbor. Here the weight is the function of trust value of each neighbor. It is defined as
\begin{equation}\label{weight}
       w_{ij}= \frac{T_{ij}}{1+\sum_{j\in N(i)}T_{ij}}.
\end{equation}
When $k \rightarrow \infty$ in (\ref{weighted_consensus}), $x_i(k)\rightarrow x^* $\cite{M05}.

The basic procedure based on (\ref{weighted_consensus}) is described as follows. At first, each virtual vehicle in the cloud service deployment, which is willing to perform cooperative spectrum sensing needs to use energy detection mechanism to explore the status of the PU. Once the status is obtained, virtual vehicles should exchange the status of the PU to each other. A virtual vehicle utilizes the sensing data received to update its value by (\ref{weighted_consensus}). When calculation finished, the virtual vehicle will disseminate its current value to its one-hop neighbors. When the difference between the new result of each virtual vehicle and a common consensus value is less than or equal to a tolerant deviation, the iterating procedure is done. Virtual vehicles compare the value with a pre-defined threshold \cite{LYH10} to determine if the PU is present.

Trust is updated for each iterating step in the process. It reflects the real trust of neighbor virtual vehicles. However, it may cause the resource limitation issues and large latency for trust retrieving. In this paper, we use the trust definition from \cite{CSC11}, which is the degree of belief that an entity (virtual vehicle) can perform a duty as expected. It is denoted as $T \in [0,1]$. Trust evaluation is employed, which is described in \cite{WYB14}.

\section{Simulation Results and Discussions}
\label{sect:simulation}
In this section, we present simulation results to show the effectiveness of the SSaaS in CR-VANETs when a malicious virtual vehicle continuously reports wrong sensing data in order to interfere the cooperative sensing. Then the probability of success is illustrated. Finally, the latency in the cloud computing is showed in the experiments.

\subsection{SSaaS Versus SSDF}
\begin{figure}[tp]
\centering{
\includegraphics[width=0.52\textwidth]{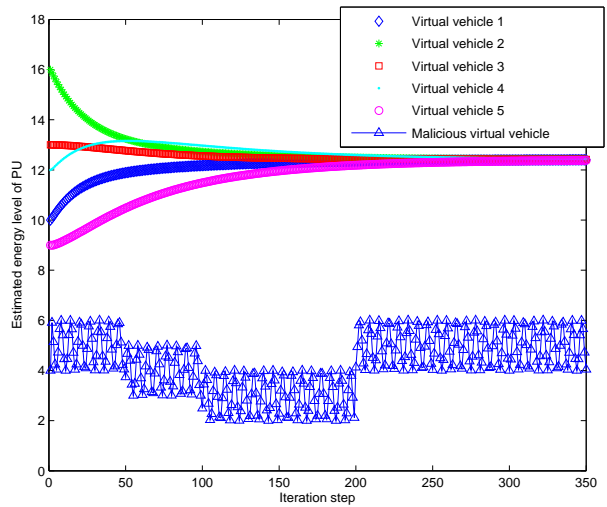}
\caption{SSaaS with one malicious virtual vehicle.}
\label{pic:dynamic_trust}}
\end{figure}

\begin{figure}[tp]
\centering{
\includegraphics[width=0.52\textwidth]{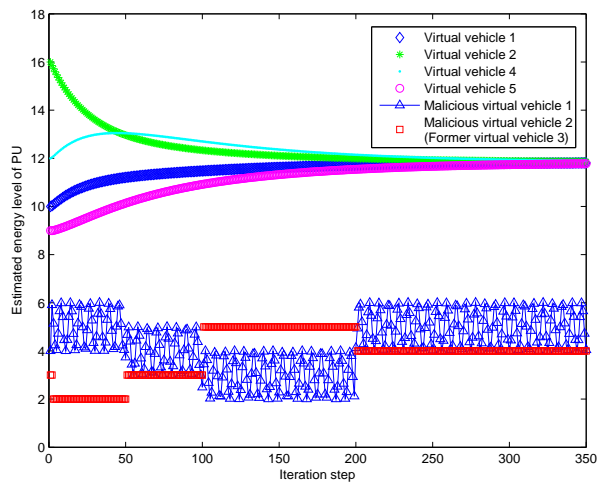}
\caption{SSaaS with two malicious virtual vehicle.}
\label{pic:dynamic_trust_2}}
\end{figure}

There are six virtual vehicles in the service deployment. There is one PU in the network. Through the cooperative spectrum sensing, the SSaaS finally provides the status of the PU. Then the user or application can decide to use the licensed spectrum band based on the result from the SSaaS.
In the simulation, we assume that the malicious virtual vehicle is always existing. The pre-defined threshold of PU presence is $\lambda = 11.4dB$. During the consensus-based cooperative spectrum sensing, the malicious virtual vehicle sends dynamic wrong sensing data to interrupt the spectrum sensing procedure. Fig. \ref{pic:dynamic_trust} shows that trust scheme can exclude the impact from the malicious virtual vehicle. The consensus result can be achieved among the normal virtual vehicles. Then the correct decision that the PU is present is made by the SSaaS. In Fig. \ref{pic:dynamic_trust_2},  virtual vehicle $3$ becomes an attacker. When two malicious virtual vehicles perform attacks with variable incorrect sensing data, the proposed service still works but the converging progress becomes slow in the simulations. In addition, simulation results show that the SSaaS is effective when the number of malicious virtual vehicles is minor in the entire virtual vehicle server list.

\subsection{Probabilities of Success}
\begin{figure}[tp]
\centering{
\includegraphics[width=0.52\textwidth]{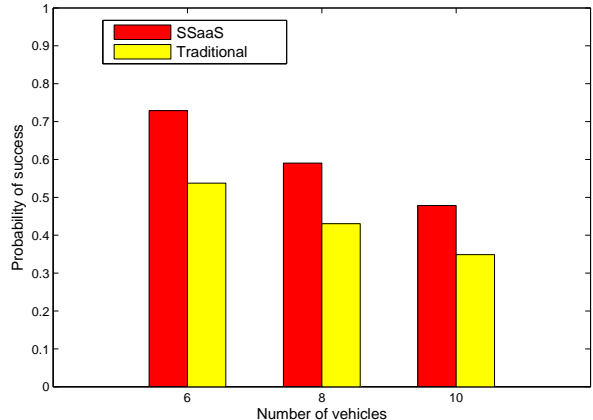}
\caption{Performance comparison in terms of the probability of success.}
\label{pic:p_s}}
\end{figure}

The benefit of the SSaaS is that the physical vehicle is virtualized in the service deployment. That means the user or application has no awareness of the real vehicles that perform spectrum sensing. The virtual vehicle can reside in the different hosts. This can utilize the resource in different hosts that have more resources, such as power, computation capability, storage etc. We compare the traditional cooperative spectrum sensing and SSaaS in terms of the probability of success, which  is defined as the probability that vehicles can perform cooperative spectrum sensing successfully and reach a finally decision, denoted as $P_s$. The availability of physical vehicles is $P_{av}$. Here, we assume $P_{av} \in [0.6,0.9]$. Fig. \ref{pic:p_s} shows that SSaaS has higher probabilities than traditional physical vehicles. This is because that virtual vehicles can be deployed to the host that has the high availability. During the spectrum sensing procedure, the VM also can be seamlessly migrated to  different hosts without interrupting the ongoing procedure thanks to the proposed lightweight cloud.

\subsection{Latency Improvement}
\begin{figure}[tp]
\centering{
\includegraphics[width=0.46\textwidth]{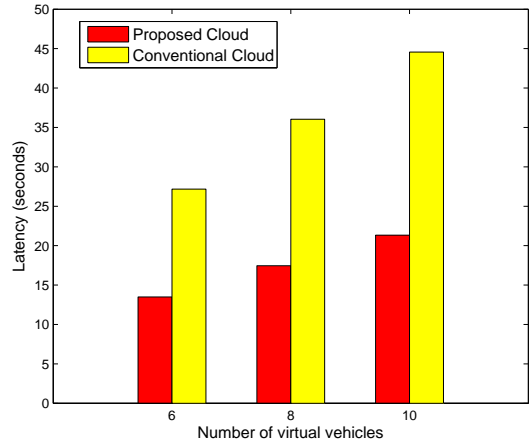}
\caption{Latency in the proposed and conventional cloud.}
\label{pic:latency}}
\end{figure}

Due to the dynamic nature of CR-VANETs, the localized cloud computing is more suitable than conventional Internet-based cloud. Here we evaluate the proposed lightweight cloud and Internet-based cloud with SSaaS service deployment in terms of number of VMs. In Fig. \ref{pic:latency}, as the number of VMs increases, the latency becomes larger gradually. This is because  each VM needs to fetch the trust from the cloud. Then the more VMs joints the cooperative spectrum sensing, the larger latency is generated. The communication time between VMs is also increased as the number of VMs increases. The latency in the conventional cloud is larger than the proposed one, almost two times. The retrieving process in the Internet-based data centers needs more time in the experiments.

\section{Conclusions and Future Work}
\label{sect:dso}
CR-VANETs have become a promising technology for driving safety and entertainment in connected vehicles. Security is the key to the success of connected vehicles. To solve the security issues of cooperative spectrum sensing in CR-VANETs, spectrum sensing as a service was proposed in this paper, which is based on the joint RSU and vehicle-based local cloud computing model. Through the deploying service in the cloud, a safe consensus-based cooperative spectrum sensing is applied. The effectiveness of this cloud-based spectrum sensing scheme is verified by the simulation results. It was shown that the cloud computing in CR-VANETs can effectively reduce latency and improve the security of CR-VANETs.
In the future work, the security issues of virtualization \cite{LY15} and cloud computing \cite{YYB15} in CR-VANETs, such as how to find the safe host during vehicular virtual machine migration \cite{BCSCADPV2016} and high availability \cite{AFGJKKLPRSZ2010}, will be studied further.

\section*{Acknowledgment}
This work was supported in part by the Natural Sciences and Engineering Research Council of Canada (NSERC).

\balance
\setlength{\baselineskip}{12pt}
\bibliographystyle{ieeetr}
\bibliography{references,D:/CA/Papers/Ref}

\end{document}